\documentclass[aip, amsmath, amssymb, reprint]{revtex4-1}  
\usepackage{graphicx}
\usepackage{dcolumn}
\usepackage{bm}
\usepackage[utf8]{inputenc}
\usepackage[T1]{fontenc}
\usepackage{mathptmx}
\usepackage{nccmath} 
\usepackage{hyperref} 
\hypersetup{colorlinks=true, citecolor=blue, urlcolor=blue, linkcolor=blue} 

\begin{document}

\title{Noise-Induced Chaos and Signal Detection by the Nonisochronous \\ Hopf Oscillator}

\author{Justin Faber}
\email{jfaber3@g.ucla.edu}
\affiliation{Department of Physics \& Astronomy, University of California, \\ Los Angeles, California 90095, USA.}
 
\author{Dolores Bozovic}
\email{bozovic@physics.ucla.edu.}
\affiliation{Department of Physics \& Astronomy, University of California, \\ Los Angeles, California 90095, USA.}
\affiliation{California NanoSystems Institute, University of California, \\ Los Angeles, California 90095, USA.}

\date{\today}

\begin{abstract}
The Hopf oscillator has been shown to capture many phenomena of the auditory and vestibular systems.  These systems exhibit remarkable temporal resolution and sensitivity to weak signals, as they are able to detect sounds that induce motion in the \AA\space regime.  In the present work, we find the analytic response function of a nonisochronous Hopf oscillator to a step stimulus and show that the system is most sensitive in the regime where noise induces chaotic dynamics.  We show that this regime also provides a faster response and enhanced temporal resolution.  Thus, the system can detect a very brief, low-amplitude pulse.  Finally, we subject the oscillator to periodic delta-function forcing, mimicking a spike train, and find the exact analytic expressions for the stroboscopic maps.  Using these maps, we find a period-doubling cascade to chaos with increasing force strength.
\end{abstract}

\maketitle

\begin{quotation}
Chaos is typically considered a harmful element in dynamical systems, as it limits their predictability and regularity.  For example, a chaotic heartbeat is an indicator of cardiac fibrillation. \cite{Garfinkel97}  Chaos may also be responsible for the anti-reliability of neurons. \cite{Goldobin06}  However, there is some evidence that the sensitivity to initial conditions that characterizes chaotic systems could be helpful for weak-signal detection.\cite{Brown92, Neiman11, Faber18b}  In the current work, we demonstrate analytically that the instabilities which give rise to chaotic dynamics in the Hopf oscillator are responsible for enhanced temporal resolution and sensitivity to weak signals.
\end{quotation}

\begin{figure*}  
\includegraphics[width=17cm]{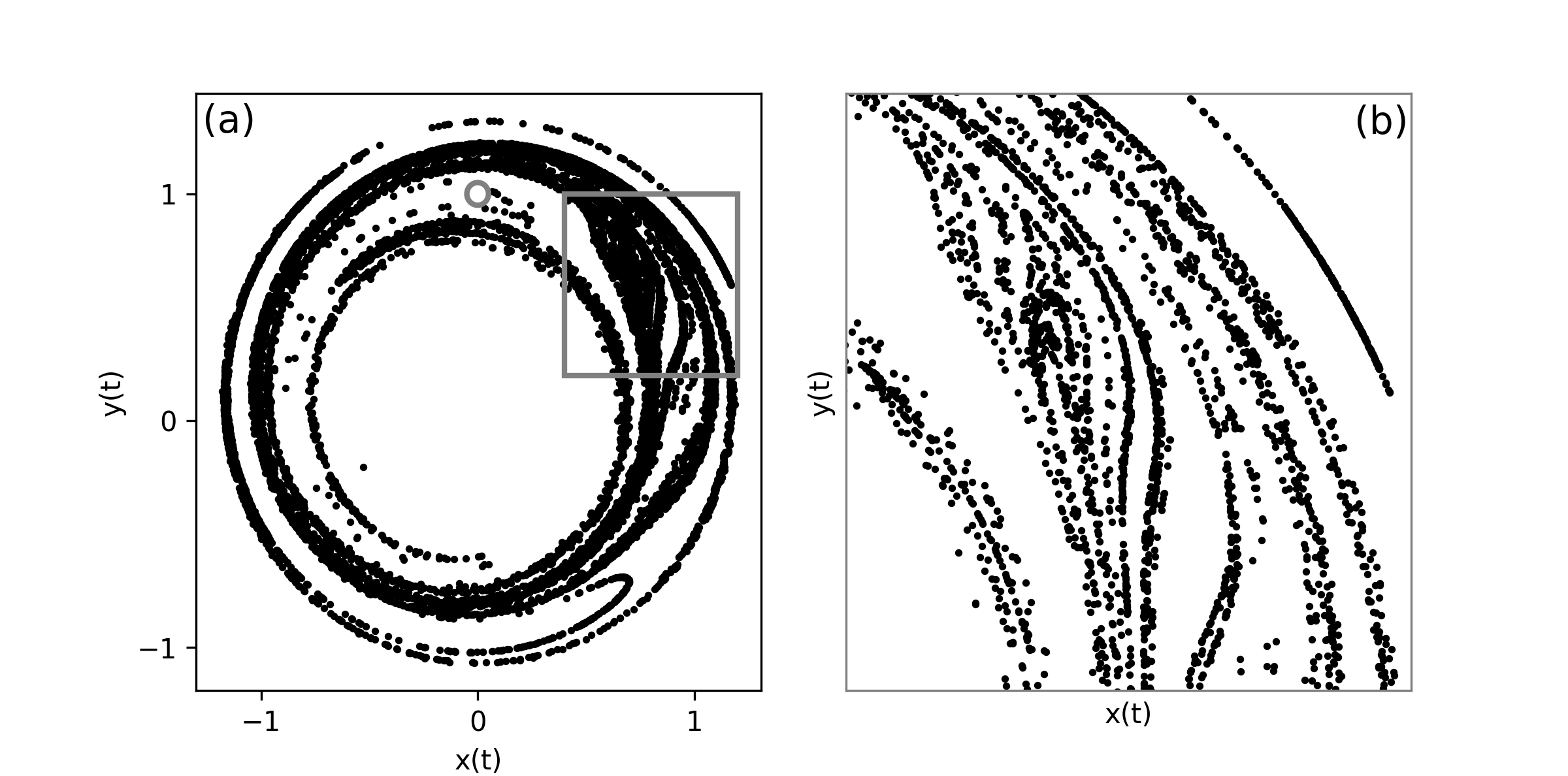}
\caption{\label{fig1} (a) The fractal structure of the Hopf oscillator with common noise:  $10^4$ slightly different initial conditions (confined to the grey circle) were evolved by 500 time steps to produce the black points.  (b) A zoomed-in version of the region in the grey box.  $\mu = \alpha = \Omega_0 = 1$,  $\beta = 100$, $D = 0.05$, where $\Omega_0$ and $D$ are the natural frequency and noise strength, respectively.}
\end{figure*}

\section{Introduction}

The auditory and vestibular systems display remarkable mechanical sensitivity.  The end organs are able to detect motion in the \AA\space regime, below the level of thermal fluctuations in the surrounding fluid. \cite{Hudspeth14}  Humans are able to resolve two clicks temporally separated by only 10 microseconds, \cite{Leshowitz71} with the stimulus waveforms presented into both ears.  The sensitivity and temporal resolution of the inner ear are not fully understood, and physics of its signal detection remains an active area of research. \cite{Reichenbach14}

Mechanical detection of sound and acceleration is performed by hair cells, sensory cell that were named after the organelle that protrudes from their apical surface.  This organelle consists of rod-like stereovilli that are organized in interconnecting rows and are collectively named the hair bundle. Transduction channels are embedded at the tips of these stereovilli and are coupled to the interconnecting tip links.  When an external signal deflects the hair bundle, it modulates the opening probability of its transduction channels. Thus, the hair bundle transduces the mechanical energy of the signal into an electrical potential change in the cell via the influx of ionic current through the channels. \cite{LeMasurier05}

Hair bundles of several species exhibit autonomous oscillations in the absence of an external stimulus. \cite{Martin03}  These spontaneous oscillations were shown to violate the fluctuation dissipation theorem, \cite{Martin01} indicating that they are a manifestation of an internal active mechanism. \cite{Hudspeth08}  We previously demonstrated in a series of experiments that these spontaneous oscillations exhibit chaotic dynamics. \cite{Faber18a}  By tuning the experimental parameters of the biological preparation, we were able to modulate the degree of chaos in the system and demonstrated that the weakly chaotic regime gives rise to enhanced sensitivity. \cite{Faber18b}.  Further, we showed that the temporal resolution of the detector increases with increasing levels of chaos.  These results were shown to be consistent with simulations of a numerical model for hair cell dynamics.

In the current work, we provide a theoretical treatment of the beneficial role of chaos in signal detection by the hair cells. We focus on noise-induced chaos, as biological sensory systems operate in thermal environments. The normal form equation for the Hopf bifurcation has  been used to model many dynamical systems. \cite{Marsden76} It was shown to capture a number of experimentally observed phenomena exhibited by the inner ear, including the sensitivity and frequency selectivity of hearing.  \cite{Eguiluz00, Kern03}  Furthermore, it provides us with the simplest model that reproduces the main characteristics of hair bundle dynamics. We consider here the nonisochronous Hopf oscillator, for which the angular frequency depends on the radius. \cite{Pikovsky01}  By varying the parameters of the system, we modulate the degree of nonisochronicity, and thus the level of chaos that would be observed in the presence of noise. \cite{Faber18b}  We calculate the sensitivity to a constant-force stimulus analytically and find the nonisochronous system to be both more sensitive and faster to respond than the isochronous one.

Experiments performed on live hair cells have shown that the bundles exhibit a non-monotonic response to large, step-function stimuli, which resembles the ringing of an underdamped oscillator. \cite{Benser96, Tinevez07}  This so-called ``twitch'' is observed only in living, active cells and is believed to be a manifestation of an internal active process, as a passive hair bundle exhibits an overdamped response. The phenomenon may play an important role in hearing, as it provides a possible mechanism by which the hair cell can amplify an incoming signal.  In the current work, we show that the Hopf oscillator can reproduce the twitch when the stimulus induces a saddle-node on invariant circle (SNIC) bifurcation resulting in a spiral-sink fixed point.  Further, we find the parameter conditions that allow such behavior, providing us with a very simple model for this biological phenomenon. 

Lastly, we explore the routes to chaos in the system, by finding the exact analytic expressions for the stroboscopic maps in the presence of a periodic delta-function stimulus.  Using these maps, we find a period-doubling cascade to chaos upon increasing the forcing strength when the system is nonisochronous.  Further, we find the same route to chaos upon increasing the degree of nonisochronicity, while keeping the forcing strength constant.

\section{Theoretical Model}

We use the normal-form equation for the Hopf bifurcation, including terms up to cubic order, and introduce additive forcing:

\begin{ceqn}
\begin{align}
\frac{dz}{dt} = (\mu + i\omega_0)z - (\alpha + i\beta)|z|^2z + F_z(t),
\label{eq:Hopf}
\end{align}
\end{ceqn}

\noindent where $z(t) = x(t) + iy(t)$ and $F_z(t) = F_x(t) + iF_y(t)$.  Here, $x(t)$ represents the bundle position, while $y(t)$ reflects internal parameters of the bundle and is not assigned a specific measurable quantity.  Chaotic dynamics arise when this system is driven by common Gaussian white noise: \cite{Zhou02}  $F_z(t) = \eta_x(t) + i\eta_y(t)$, where $\langle \eta_x(t)\eta_x(t') \rangle  = \langle \eta_y(t)\eta_y(t') \rangle = 2D \delta (t-t')$, and $\langle \eta_x(t)\eta_y(t') \rangle  = 0$ (Fig. \ref{fig1}).  We previously demonstrated that, under these conditions, the Lyapunov exponent can be approximated as \cite{Faber18b}

\begin{ceqn}
\begin{align}
\lambda \approx \frac{D\beta}{\mu}.
\label{eq:Lyapunov}
\end{align}
\end{ceqn}

\noindent When $\beta \neq 0$, the oscillator is nonisochronous: the frequency is dependent on the amplitude of oscillations.  Prior work has demonstrated that nonisochronous systems exhibit noise-induced chaos \cite{Faber18b}.  For simplicity, we study this instability in a noiseless system, while noting that in the presence of noise, the degree of chaos would be proportional to $\beta$.

Auditory and vestibular stimuli induce lateral deflections on the hair bundle, so we consider forces in the $\hat{x}$ direction, which coincides with the direction of autonomous oscillation.  In polar coordinates, the model takes the form:

\begin{ceqn}
\begin{align}
\frac{dr}{dt} = \mu r - \alpha r^3 + F_x(t)\cos\theta
\label{eq:r_dot}
\end{align}
\end{ceqn}

\begin{ceqn}
\begin{align}
\frac{d\theta}{dt} = \omega_0 - \beta r^2 - F_x(t)\frac{\sin\theta}{r}.
\label{eq:theta_dot}
\end{align}
\end{ceqn}

\noindent We observe that, in the absence of forcing, there is a stable limit cycle of radius $r_0 = \sqrt{\frac{\mu}{\alpha}}$ for $\mu > 0$.  Further, there is a nullcline in the phase at radius $r_{nc} = \sqrt{\frac{\omega_0}{\beta}}$.  If the force causes the stable limit cycle to approach this radius, the angular frequency slows down, and the system approaches a SNIC bifurcation.  We show that the forcing can perturb the shape of the limit cycle, causing an intersection between the stable limit cycle and $r_{nc}$.  This SNIC bifurcation causes a pair of fixed points (one stable, one unstable) to appear on the limit cycle (see Appendix A).  When the system is poised in close proximity to this bifurcation, noise can most easily induce a chaotic response, and the system is very sensitive to external signals.

\section{Response to a Constant-Force Stimulus}

We consider forcing of the form $F_x(t) = f_0 \Theta(t-t_0)$, where $\Theta$ is the Heaviside step function, and look at the response of a system in the oscillatory regime ($\mu > 0$).  We assume that the forcing is weak compared to the stability of the limit cycle ($\frac{f_0}{\mu r_0} << 1$).  We can therefore assume that the forcing acts as a small perturbation on the shape of the limit cycle and ignore the transient in $r$ at the onset of the forcing.  Looking for solutions near the limit cycle, we let $r(t) = r_0 + \delta r(t)$, where $r_0 = \sqrt{\frac{\mu}{\alpha}}$ is the limit cycle radius in the absence of forcing.  Inserting this approximation into Eq. (\ref{eq:r_dot}) and keeping only terms that are linear in $\delta r$ yields

\begin{ceqn}
\begin{align}
\frac{d}{dt}\delta r = -2\mu\delta r + f_0 \Theta(t-t_0)\cos\theta,
\end{align}
\end{ceqn}

\noindent which has a steady-state solution $\delta r = \frac{f_0\cos\theta}{2\mu}$ for $t > t_0$.  Therefore, the first order perturbation to the shape of the limit cycle is given by:

\begin{ceqn}
\begin{align}
r(t) = r_0 + \frac{f_0\cos\theta(t)}{2\mu}.
\label{eq:r}
\end{align}
\end{ceqn}

\noindent Inserting this solution into Eq. (\ref{eq:theta_dot}) and keeping only the first-order forcing terms, we find that

\begin{ceqn}
\begin{align}
\frac{d\theta}{dt} = \Omega_0 - \frac{f_0\sin\theta}{r_0} - \frac{\beta r_0 f_0\cos\theta}{\mu},
\end{align}
\end{ceqn}

\noindent where $\Omega_0 = \omega_0 - \beta r_0^2$ is the natural frequency.  We integrate the equation to solve for $\theta (t)$:

\begin{ceqn}
\begin{align}
\int_{t_0}^{t} dt' = \int_{\theta_0}^{\theta} \frac{d\theta'}{\Omega_0 - \frac{f_0\sin\theta'}{r_0} - \frac{\beta r_0 f_0\cos\theta'}{\mu}},
\end{align}
\end{ceqn}

\noindent where $\theta_0$ is the phase of the oscillator at the onset of the step. Evaluating the integral yields

\begin{ceqn}
\begin{align}
-\frac{1}{2} \gamma (t - t_0)  =  \tanh^{-1}(Q(\theta)) - \tanh^{-1}(Q(\theta_0)),
\label{eq:tanh}
\end{align}
\end{ceqn}

\noindent where 

\begin{ceqn}
\begin{align}
\gamma = \sqrt{ \Big(\frac{f_0}{r_0}\Big)^2 + \Big(\frac{\beta r_0 f_0}{\mu}\Big)^2 - \Omega_0^2}
\label{eq:gamma} 
\end{align}
\end{ceqn}

\noindent and 

\begin{ceqn}
\begin{align}
Q(\theta) = \frac{(\Omega_0 + \frac{\beta r_0 f_0}{\mu})\tan(\frac{\theta}{2}) - \frac{f_0}{r_0}}{\gamma}.
\end{align}
\end{ceqn}

\noindent Inverting Eq. (\ref{eq:tanh}) yields

\begin{ceqn}
\begin{align}
\theta(t) = 2 \tan^{-1} \Bigg( \frac{ \gamma \tanh \Big(-\frac{1}{2} \gamma (t - t_0) + \tanh^{-1}(Q(\theta_0))\Big) + \frac{f_0}{r_0} }{ \Omega_0 + \frac{\beta r_0 f_0}{\mu} } \Bigg).
\label{eq:theta}
\end{align}
\end{ceqn}

\noindent Using the identity $\tanh^{-1}(x) = \frac{1}{2}\log(1 + x) - \frac{1}{2}\log(1 - x)$, we can also write Eq. (\ref{eq:tanh}) in exponental form:

\begin{ceqn}
\begin{align}
e^{- \gamma (t - t_0)}  =  \frac{(1 + Q(\theta))(1 - Q(\theta_0))}{(1 - Q(\theta))(1 + Q(\theta_0))}.
\label{eq:exp}
\end{align}
\end{ceqn}

\begin{figure*}[t!]  
\includegraphics[width=17cm]{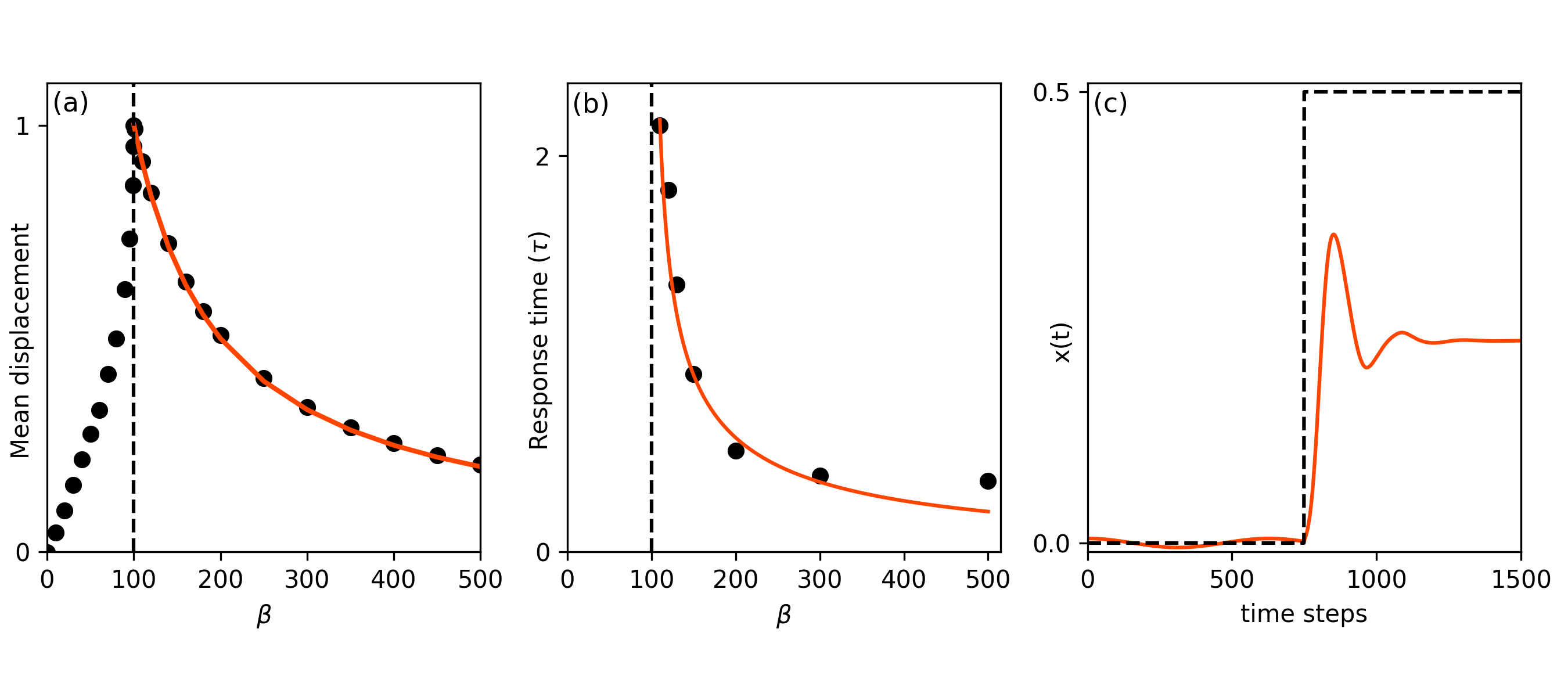}
\caption{\label{fig2} (a) Mean displacement of the Hopf oscillator in response to a constant-force stimulus found semi-analytically (black points) and fully analytically (solid curve).  (b) Response time of the oscillator characterized by the exponential decay time in response to a step force.  Black points represent the values found from numerical simulations, and the solid curve represents the analytic approximation.  For (a-b),  $\mu = \alpha = \Omega_0 = 1$ and $f_0 = 0.01$.  The dashed line indicates the point where the forcing is able to induce a SNIC bifurcation.  (c) An example of the twitch, a non-monotonic response (solid) to a step force (dashed), for $\mu = \alpha = \Omega_0 = 1$, $\beta = 50$, and $f_0 = 0.1$.  The response curve represents the average over 200 different initial conditions, evenly separated in phase.}
\end{figure*}

\subsection{Response Amplitude}

We use the mean displacement in $x$ to characterize the response amplitude:

\begin{ceqn}
\begin{align} 
\langle \Delta x(t) \rangle = \langle r(t)\cos\theta(t) \rangle_{stim} - \langle r(t)\cos\theta(t)\rangle_{0},
\end{align}
\end{ceqn}

\noindent where the first and second terms on the right side are time averages in the presence and absence of the stimulus, respectively.  The second term yields zero.  Inserting Eq. (\ref{eq:r}), we obtain

\begin{ceqn}
\begin{align} 
\langle \Delta x(t) \rangle = r_0 \langle \cos\theta(t) \rangle_{stim} + \frac{f_0}{2\mu} \langle \cos^{2}\theta(t) \rangle_{stim},
\label{eq:mean_disp}
\end{align}
\end{ceqn}

\noindent which can be solved semi-analytically by inserting Eq. (\ref{eq:theta}) and taking the time average over a long time series (Fig. \ref{fig2}).

Depending on the choice of parameters, $\gamma$ is either purely real or purely imaginary. Eq. (\ref{eq:exp}) indicates that the solutions are oscillatory for imaginary $\gamma$ and decay exponentially to a fixed point for real $\gamma$.  In the latter case, we can find the mean displacement analytically.  This fixed point corresponds to the intersection between the perturbed limit cycle, $r(t) = r_0 + \frac{f_0\cos\theta}{2\mu}$, and the angular nullcline $r_{nc} = \sqrt{\frac{\omega_0}{\beta}}$.  We set these two radii equal to each other and find that

\begin{ceqn}
\begin{align} 
\cos\theta = \frac{2\mu}{f_0} \bigg(\sqrt{\frac{\omega_0}{\beta}} - r_0 \bigg).
\end{align}
\end{ceqn}

\noindent We insert this solution into Eq. (\ref{eq:mean_disp}) and substitute in the natural frequency, $\Omega_0 = \omega_0 - \beta r_0^2$, to obtain:

\begin{ceqn}
\begin{align} 
\langle \Delta x(t) \rangle = \frac{2\mu r_0}{f_0} \bigg(\sqrt{\frac{\Omega_0}{\beta} + r_0^2} - r_0 \bigg) +  \frac{2\mu}{f_0} \bigg(\sqrt{\frac{\Omega_0}{\beta} + r_0^2} - r_0 \bigg)^2.
\label{mean_disp_analytic}
\end{align}
\end{ceqn}

\noindent This analytic curve is plotted in Fig. \ref{fig2}.  Note that at a fixed natural frequency $\Omega_0$ and in the limit of large $\beta$, the radius of the angular nullcline $r_{nc}$ approaches the limit cycle radius.  With these two circles close together, any small perturbation to the limit cycle can cause an intersection, resulting in a stable fixed point.

\subsection{Temporal Resolution}

We characterize the temporal resolution of the system by calculating the response time (decay time) to a step stimulus.  Our results are limited to real $\gamma$, as this regime shows an exponential decay to a stable fixed point upon the application of a stimulus.  For imaginary $\gamma$, the system would continue to oscillate in the presence of a step stimulus.  We perform a Taylor series expansion of the right side of Eq. (\ref{eq:exp}) in the vicinity of the stable fixed point (see Appendix B):

\begin{ceqn}
\begin{align}
\theta(t) \approx \frac{1-Q(\theta_c)^2}{2Q'(\theta_c)}\Big(e^{- \gamma (t - t_0)} - 1\Big) + \theta_0,
\end{align}
\end{ceqn}

\noindent where $\theta_c$ is the angle of the stable fixed point.  The decay time of this solution is

\begin{ceqn}
\begin{align}
\tau_{res} = \frac{1}{\gamma} = \frac{1}{ \sqrt{ (\frac{f_0}{r_0})^2 + (\frac{\beta r_0 f_0}{\mu})^2 - \Omega_0^2} }.
\end{align}
\end{ceqn}

\noindent We fit exponential functions to the numerical simulations of the response and compare the values to the analytic form of $\tau_{res}$ in Fig. \ref{fig2}.

\begin{figure*}[t!]  
\includegraphics[width=17cm]{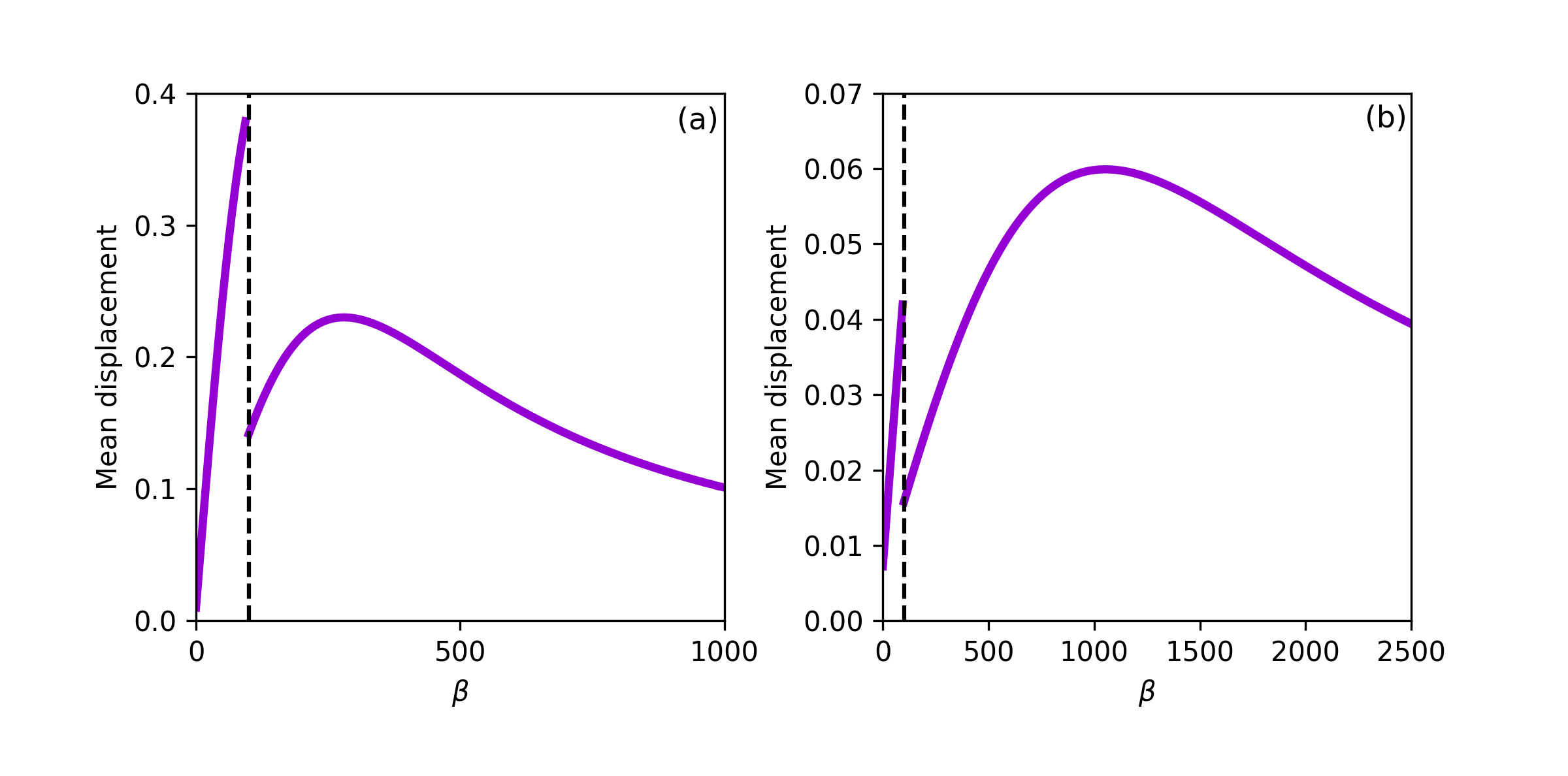}
\caption{\label{fig3} Mean displacement of the Hopf oscillator in response to a brief, constant-force stimulus.  (a) and (b) represent the responses to a pulse of duration $\frac{1}{4}$ and $\frac{1}{16}$ of the natural period of oscillation, respectively.  $\mu = \alpha = \Omega_0 = 1$, $f_0 = 0.01$, and the vertical dashed line represents the value of $\beta$ at which the forcing is able to induce a SNIC bifurcation.}
\end{figure*}

\subsection{Non-monotonic response}

In the regime of large $\beta$ and/or strong forcing, the averaged response curve does not resemble an exponential decay.  Rather, the response is non-monotonic and resembles the observed phenomenon known in hair bundle mechanics as ``the twitch'' \cite{Benser96, Tinevez07} (Fig. \ref{fig2}). In our simple model, the non-monotonic behavior is a result of a spiral sink forming after the step force induces a SNIC bifurcation.  The trajectories spiraling to the fixed point give rise to the ringing behavior in $x(t)$.  In this section, we find the conditions necessary for the Hopf oscillator to produce this phenomenon.  We find the fixed point $(R_{c}, \theta_{c})$ using Eqs. (\ref{eq:r_dot}) and (\ref{eq:theta_dot}):

\begin{ceqn}
\begin{align}
0 = \mu R_{c} - \alpha R_{c}^3 + f_0\cos\theta_{c}
\label{eq:r_dot_fp}
\end{align}
\end{ceqn}

\begin{ceqn}
\begin{align}
0 = \omega_0 - \beta R_{c}^2 - f_0\frac{\sin\theta_{c}}{R_{c}}.
\label{eq:theta_dot_fp}
\end{align}
\end{ceqn}

\noindent The Jacobian at the fixed point takes the form:

\begin{ceqn}
\begin{align}
J = 
\quad
\begin{pmatrix} 
\mu - 3\alpha R_{c}^2 & -f_0\sin\theta_{c} \\
- 2\beta R_{c} + f_0\frac{\sin\theta_{c}}{R_{c}^2} & - f_0\frac{\cos\theta_{c}}{R_{c}}
\end{pmatrix}
\quad 
\end{align}
\end{ceqn}

\noindent which can also be written as

\begin{ceqn}
\begin{align}
J = 
\quad
\begin{pmatrix} 
\mu - 3\alpha R_{c}^2 & \beta R_c^3 - \omega_0 R_c \\
 \frac{\omega_0}{R_c} - 3\beta R_c & \mu - \alpha R_c^2
\end{pmatrix}
\quad 
\end{align}
\end{ceqn}

\noindent by using Eqs. (\ref{eq:r_dot_fp}) and (\ref{eq:theta_dot_fp}) to eliminate $\theta_c$.  A spiral sink is present if and only if the trace of J is less than zero, the determinant of J is greater than zero, and $4\det(J) - Tr(J)^2 > 0$. \cite{Strogatz94}  Since we are considering the limit cycle regime, we have $\mu > 0$.  We will also use the condition that $R_c \geq r_0 = \sqrt{\frac{\mu}{\alpha}}$.  Since this effect is most pronounced when $\beta$ is large, we will assume that the second term in Eq. (\ref{eq:theta_dot}) is larger in magnitude than the first ($\beta r_0^2 > \omega_0$).  With these assumptions, the first two conditions are always satisfied:

\begin{ceqn}
\begin{align}
Tr(J) = 2\mu - 4\alpha R_c^2 \leq 2\mu - 4\alpha r_0^2 = -2\mu < 0
\end{align}
\end{ceqn}

\noindent and

\begin{eqnarray}
\begin{split}
\det(J) = \mu^2 \bigg(3\Big(\frac{R_c}{r_0}\Big)^4 - 4\Big(\frac{R_c}{r_0}\Big)^2 + 1 \bigg)   \\
+ \omega_0^2 \bigg( 3\Big(\frac{\beta R_c^2}{\omega_0}\Big)^2 - 4\Big(\frac{\beta R_c^2}{\omega_0}\Big) + 1 \bigg)  >  0,
\end{split}
\end{eqnarray}

\noindent since $\frac{R_c}{r_0} \geq 1$ and $\frac{\beta R_c^2}{\omega_0} \geq \frac{\beta r_0^2}{\omega_0} > 1$.  The third and final condition is satisfied when

\begin{ceqn}
\begin{align}
\beta > \frac{2\omega_0 + \sqrt{3\mu^2 + \omega_0^2}}{3r_0^2}
\end{align}
\end{ceqn}

\noindent  and $\beta > \frac{\alpha}{\sqrt{3}}$ (see Appendix C).  Thus, we have found the conditions on the parameters of the Hopf oscillator such that it can reproduce the experimentally observed twitch.

\section{Response to a Short-Pulse Stimulus}

We now find the response amplitude to a brief, weak force of amplitude $f_0$ and duration $T$:

\begin{ceqn}
\begin{align}
F_x(t) = f_0 \Big(\Theta(t) - \Theta(t - T)\Big).
\end{align}
\end{ceqn}

\noindent The solution during the force step was already found in Eqs. (\ref{eq:r}) and (\ref{eq:theta}), provided that the initial conditions at the onset of the pulse are on the stable limit cycle.  We use these equations to find the displacement induced in $x(t)$.  The mean displacement is obtained by computing the average of $(x(T) - x(0))$ over many different initial conditions.  The mean displacement is used to quantify the responsiveness of the system.

In Fig. \ref{fig3}, we show that there is a local maximum in responsiveness at the SNIC bifurcation, as in the steady-state case.  However, we observe an additional local maximum as $\beta$ increases beyond the bifurcation point.  Increasing $\beta$ in this regime increases the speed at which the trajectories move across the circle to the fixed point.  This increase is beneficial for the responsiveness of the system.  However, the increase in $\beta$  also moves the fixed point in the $-\hat{x}$ direction, thereby reducing the mean displacement in $x$ (see Appendix A).  The competition between these two effects results in a local maximum.  The location of this maximum is dependent on the pulse duration, which determines the time needed for the trajectories to reach the fixed point, starting from the initial conditions.

\begin{figure*}[t!]  
\includegraphics[width=17cm]{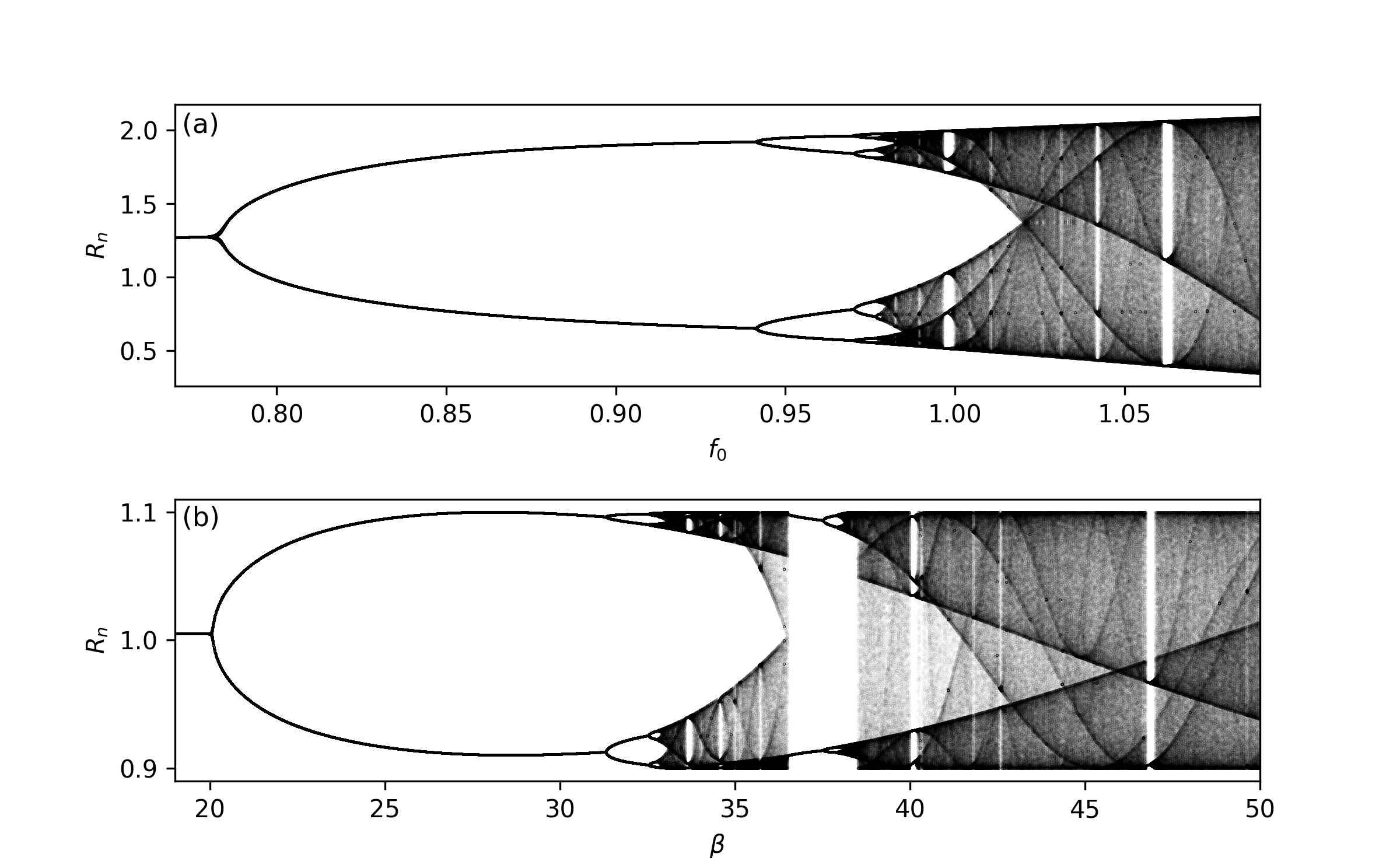}
\caption{\label{fig4}  Bifurcation diagrams generated from the radial stroboscopic map at each impulse of the spike train.  The interval between spikes was set to be the natural period of oscillations ($\tau = \frac{2\pi}{\Omega_0}$, $\mu = \alpha = \Omega_0 = 1$).  (a) $\beta = 3$, while $f_0$ is varied.  (b) $f_0 = 0.1$, while $\beta$ is varied.}  
\end{figure*}

\section{Effects of a Spike Train}

We now find the response to periodic, delta-function forcing.  We construct the stroboscopic Poincar\'e maps analytically \cite{Gonzalez83} and observe the transition to chaos.  First, we find the analytic solution in the absence of forcing by integrating Eqs. (\ref{eq:r_dot}) and (\ref{eq:theta_dot}) with $F_x(t) = 0$.  The first can be solved using partial fraction decomposition:

\begin{ceqn}
\begin{align}
\int_{t_i}^t dt' = \int_{r_i}^r \frac{dr'}{\mu r' - \alpha r'^3}  = \int_{r_i}^r  \Big( \frac{1}{\mu r'} - \frac{\alpha r'}{\mu (\alpha r'^2 - \mu)} \Big)dr',
\end{align}
\end{ceqn}

\noindent which yields

\begin{ceqn}
\begin{align}
r_{so}(t, r_i) = \sqrt{ \frac{\mu}{ \alpha - (\alpha - \frac{\mu}{r_i^2})e^{-2\mu t} } },
\end{align}
\end{ceqn}

\noindent where we have set the initial time $t_i$ to zero, and $r_i$ is the radius at this time.  We insert this solution into Eq. (\ref{eq:theta_dot}):

\begin{ceqn}
\begin{align}
\frac{d\theta}{dt} = \omega_0 - \frac{\beta\mu}{ \alpha - (\alpha - \frac{\mu}{r_i^2})e^{-2\mu t} }.
\end{align}
\end{ceqn}

\noindent Integrating this equation yields

\begin{ceqn}
\begin{align}
\theta_{so}(t, r_i, \theta_i) = \theta_i + \omega_0 t + \frac{\beta}{2\alpha} \log \Bigg( \frac{\mu}{\alpha r_i^2 (e^{2\mu t} - 1) + \mu}  \Bigg), 
\end{align}
\end{ceqn}

\noindent where $\theta_i$ is the phase at $t=0$.  We now include the spike-train stimulus:

\begin{ceqn}
\begin{align}
F_x(t) = f_0 \sum_{n=1}^{n=\infty} \delta(t - n\tau).
\end{align}
\end{ceqn}

\noindent The first impulse occurs at time $t=\tau$.  Before this time ($t = \tau - \epsilon$), the solution is simply $r(\tau - \epsilon) = r_{so}(\tau - \epsilon, r_i)$ and $\theta(\tau - \epsilon) = \theta_{so}(\tau - \epsilon, r_i, \theta_i)$.  Projecting $F_x(\tau)$ onto polar coordinates, we find the solution at the time of the first impulse to be:

\begin{ceqn}
\begin{align}
r(t = \tau) = R_1 = r_{so}(\tau, r_i) + f_0 \cos\big( \theta_{so}(\tau, r_i, \theta_i)\big)
\end{align}
\end{ceqn}

\begin{ceqn}
\begin{align}
\theta(t=\tau) = \Theta_1 = \theta_{so}(\tau, r_i, \theta_i)  -  \frac{f_0 \sin\big( \theta_{so}(\tau, r_i, \theta_i)\big)}{ r_{so}(\tau, r_i) }.
\end{align}
\end{ceqn}

\noindent The solution between the first and second impulses ($t = \tau + \epsilon$) takes the form:

\begin{ceqn}
\begin{align}
r(\tau + \epsilon) = r_{so}(\epsilon, R_1)
\end{align}
\end{ceqn}

\begin{ceqn}
\begin{align}
\theta(\tau + \epsilon) = \theta_{so}(\epsilon, R_1, \Theta_1),
\end{align}
\end{ceqn}

\noindent where we treat the coordinates at the previous impulse as the initial conditions and reset the time.  We now construct the stroboscopic Poincar\'e maps using this recursive method.  We find the radius and phase at every spike, $R_n = r(t=n\tau)$ and $\Theta_n = \theta(t=n\tau)$, to be:

\begin{ceqn}
\begin{align}
R_{n+1} = r_{so}(\tau, R_n) + f_0 \cos\big( \theta_{so}(\tau, R_n, \Theta_n)\big)
\end{align}
\end{ceqn}

\begin{ceqn}
\begin{align}
\Theta_{n+1} = \theta_{so}(\tau, R_n, \Theta_n)  -  \frac{f_0 \sin\big( \theta_{so}(\tau, R_n, \Theta_n)\big)}{ r_{so}(\tau, R_n) }.
\end{align}
\end{ceqn}

\noindent These stroboscopic maps are valid for both the oscillatory and quiescent regimes.  We have made no assumptions on the choice of parameters or the strength of the forcing.  Upon increasing either $f_0$ or $\beta$, we observe a period-doubling cascade to chaos (Fig. \ref{fig4}).

\section{Discussion}

In a prior study, we demonstrated that nonisochronicity results in noise-induced chaos for even very weak levels of noise and any nonzero $\beta$.  In the present work, we demonstrated analytically that nonisochronicity is responsible for enhanced responsiveness to weak force steps, with the most responsive system being poised near the SNIC bifurcation. Further, we demonstrated that detection of a single constant-force pulse is highly rapid in a nonisochronous system. By varying $\beta$ in the Hopf oscillator equation, we control the degree of nonisochronicity and study its impact on response characteristics. We show that the oscillator can detect very short pulses for a wide range of $\beta$, with the ideal value depending on the duration of the pulse.  We also showed that the speed of the response, as measured by the exponential decay time to the constant force, increases with increasing $\beta$.  

Although we considered the noiseless system in this work, upon including even a small amount of noise, $\beta$ can be directly mapped onto the Lyapunov exponent. We previously demonstrated enhancement of sensitivity and temporal resolution with numerical models, where we varied the Lyapunov exponent directly \cite{Faber18b}.  The results of this numerical study are consistent with that of the present work. Nonisochronicity in a noiseless system hence directly correlates with the degree of chaos in the presence of noise. 

Numerical models of hair cell dynamics typically require many variables to reproduce the experimentally observed twitch.  Here, we demonstrated that this 2-dimensional model can capture the main features of this effect.  This has, thus far, been the simplest model of hair cell dynamics able to produce this behavior.  We have shown that the twitch can be explained through bifurcation theory: the stimulus yields a SNIC bifurcation with one of the fixed points being a spiral sink.  Further, we found the parameter conditions required to see the effect. 

Finally, we found the exact analytic expressions for the stroboscopic Poincar\'e maps in $r$ and $\theta$, when the oscillator is subjected to periodic, impulsive forcing.  These maps are valid for any parameter regime and any degree of forcing.  Using these stroboscopic maps, we observe a period-doubling cascade to chaos upon increasing either the forcing strength or $\beta$.  The nature of this stimulus resembles that of efferent activity.  In hair cell dynamics, efferent activity is believed to be the biological feedback responsible for modulating the sensitivity of the detector and may serve as a protective mechanism in preventing damage caused by acoustic trauma. \cite{Patuzzi91}  In the present work, we demonstrated that an efferent-like stimulus can induce or modulate the degree of chaos, thus affecting the sensitivity of the system.  Future work entails measuring the change in chaoticity and sensitivity caused by efferent activity.

Chaos is often considered an unfavorable element in dynamical systems, as it limits their predictability and regularity.  However, chaos is likely present in a number of real-world systems, as they contain many degrees of freedom and exhibit nonlinearities. Here, we have shown analytically that the instabilities that give rise to chaotic dynamics are also responsible for enhanced sensitivity and temporal resolution in the Hopf oscillator, suggesting a beneficial role of chaos in the auditory and vestibular systems.  We propose that chaos may play a role in signal detection in other noisy biological systems where timing and sensitivity are essential.

\begin{acknowledgments}
The authors gratefully acknowledge support of NSF Physics of Living Systems, under grant 1705139.
\end{acknowledgments}

\appendix  

\section{SNIC Bifurcation}

\begin{figure}[h!]
\includegraphics[width=8.5cm]{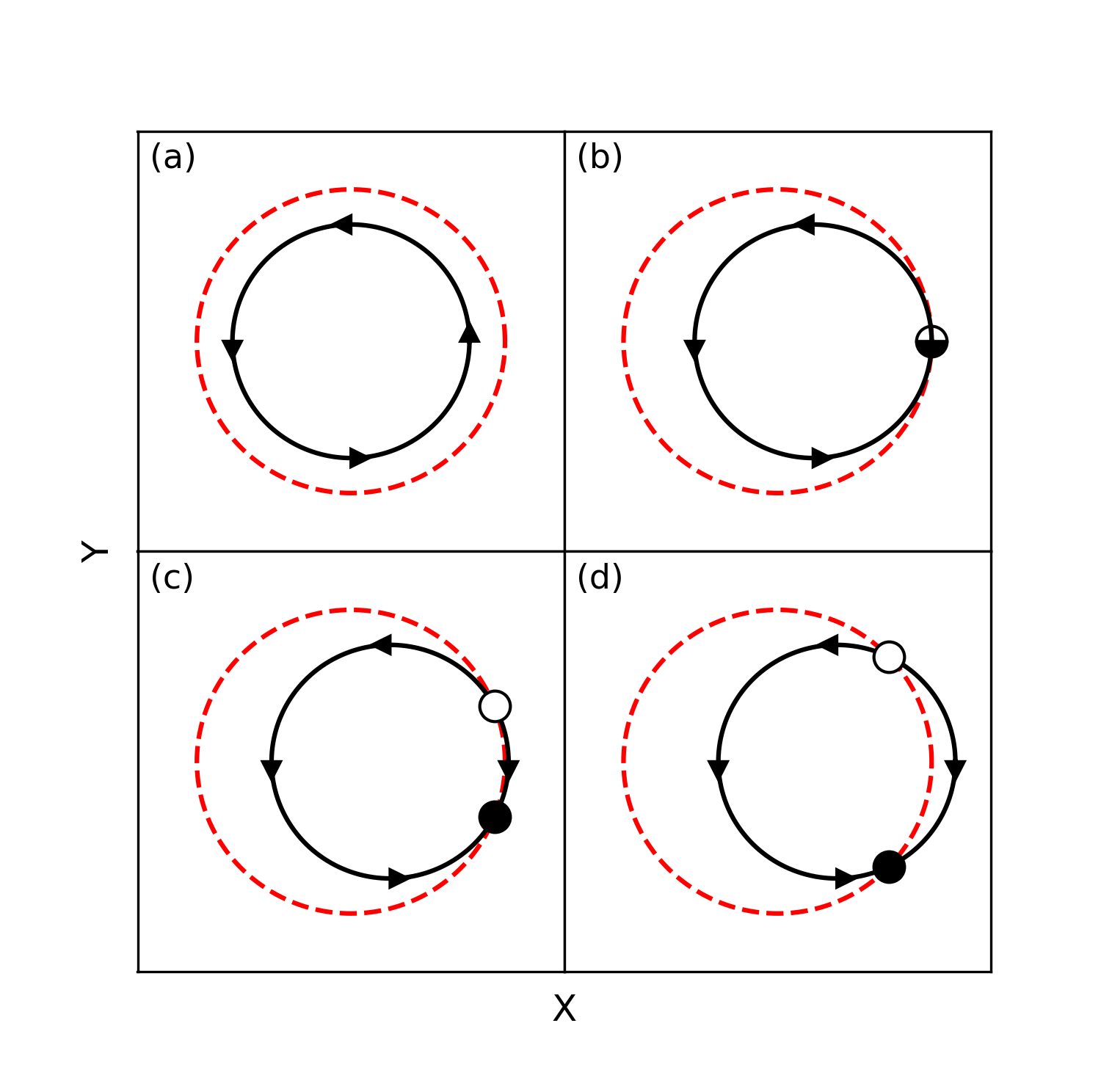}
\caption{\label{fig5}  Illustration of the SNIC bifurcation.  The solid and dashed curves represent the stable limit cycle and angular nullcline, respectively.  Filled and empty points represent attracting and repelling fixed points, respectively.  (a) Phase space diagram with no forcing.  (b) A constant force in the $\hat{x}$ direction pushes the limit cycle into the angular nullcline, producing a semi-stable fixed point.  (c) Upon increasing the force, the semi-stable fixed point splits into an attracting and a repelling fixed point.  (d) Stronger forcing causes these two points to separate further.}  
\end{figure}

\section{Response Time}

We perform a Taylor series expansion of the right side of Eq. (\ref{eq:exp}) around $\theta_c$, the angle of the stable fixed point.  Keeping only first-order terms $(Q(\theta) \approx Q(\theta_c) + Q'(\theta_c)(\theta - \theta_c))$ yields

\begin{eqnarray}
\begin{split}
e^{- \gamma (t - t_0)} \approx 
\frac{(1+Q(\theta_c)+Q'(\theta_c)(\theta - \theta_c))}
{(1-Q(\theta_c)-Q'(\theta_c)(\theta - \theta_c))}  \\
\times \frac{(1-Q(\theta_c)-Q'(\theta_c)(\theta_0 - \theta_c))}{(1+Q(\theta_c)+Q'(\theta_c)(\theta_0 - \theta_c))}.
\end{split}
\end{eqnarray}

\noindent Expanding and keeping only linear terms in $(\theta - \theta_c)$ and $(\theta_0 - \theta_c)$ results in

\begin{ceqn}
\begin{align}
e^{- \gamma (t - t_0)} \approx 1 + \frac{2Q'(\theta_c)}{1-Q(\theta_c)^2} (\theta - \theta_0)
\end{align}
\end{ceqn}

\noindent or

\begin{ceqn}
\begin{align}
\theta(t) \approx \frac{1-Q(\theta_c)^2}{2Q'(\theta_c)}\Big(e^{- \gamma (t - t_0)} - 1\Big) + \theta_0.
\end{align}
\end{ceqn}

\noindent This solution has response time 

\begin{ceqn}
\begin{align}
\tau_{res} = \frac{1}{\gamma} = \frac{1}{ \sqrt{ (\frac{f_0}{r_0})^2 + (\frac{\beta r_0 f_0}{\mu})^2 - \Omega_0^2} }.
\end{align}
\end{ceqn}

\section{Parameter Conditions for the Twitch}

The third condition required for a spiral sink is $4\det(J) - Tr(J)^2 > 0$, yielding

\begin{eqnarray}
\begin{split}
4\mu^2 \bigg(3\Big(\frac{R_c}{r_0}\Big)^4 - 4\Big(\frac{R_c}{r_0}\Big)^2 + 1 \bigg)   \\
+ 4\omega_0^2 \bigg( 3\Big(\frac{\beta R_c^2}{\omega_0}\Big)^2 - 4\Big(\frac{\beta R_c^2}{\omega_0}\Big) + 1 \bigg)   \\
- 4\mu^2 \bigg( 4\Big(\frac{R_c}{r_0}\Big)^4 - 4\Big(\frac{R_c}{r_0}\Big)^2 + 1 \bigg)  > 0,
\end{split}
\end{eqnarray}

\noindent which simplifies to

\begin{ceqn}
\begin{align}
\omega_0^2 \bigg( 3\Big(\frac{\beta R_c^2}{\omega_0}\Big)^2 - 4\Big(\frac{\beta R_c^2}{\omega_0}\Big) + 1 \bigg)  - \frac{\mu^2 R_c^4}{r_0^4} > 0.
\end{align}
\end{ceqn}

\noindent We can rewrite this condition as

\begin{ceqn}
\begin{align}
(3\beta^2 r_0^4 - \mu^2) \Big(\frac{R_c}{r_0}\Big)^4 - 4\beta\omega_0 r_0^2 \Big(\frac{R_c}{r_0}\Big)^2 + \omega_0^2 > 0,
\end{align}
\end{ceqn}

\noindent which is always satisfied if the leading order term is greater than zero $(\beta > \frac{\alpha}{\sqrt{3}})$ and 

\begin{ceqn}
\begin{align}
3\beta^2 r_0^4 - \mu^2 - 4\beta\omega_0 r_0^2 + \omega_0^2 > 0,
\end{align}
\end{ceqn}

\noindent since $\frac{R_c}{r_0} \geq 1$.  Solving for $\beta$ yields

\begin{ceqn}
\begin{align}
\beta > \frac{2\omega_0 + \sqrt{3\mu^2 + \omega_0^2}}{3r_0^2}.
\end{align}
\end{ceqn}

\bibliography{Bibliography}

\providecommand{\noopsort}[1]{}\providecommand{\singleletter}[1]{#1}%
\begin{thebibliography}{23}%
\makeatletter
\providecommand \@ifxundefined [1]{%
 \@ifx{#1\undefined}
}%
\providecommand \@ifnum [1]{%
 \ifnum #1\expandafter \@firstoftwo
 \else \expandafter \@secondoftwo
 \fi
}%
\providecommand \@ifx [1]{%
 \ifx #1\expandafter \@firstoftwo
 \else \expandafter \@secondoftwo
 \fi
}%
\providecommand \natexlab [1]{#1}%
\providecommand \enquote  [1]{``#1''}%
\providecommand \bibnamefont  [1]{#1}%
\providecommand \bibfnamefont [1]{#1}%
\providecommand \citenamefont [1]{#1}%
\providecommand \href@noop [0]{\@secondoftwo}%
\providecommand \href [0]{\begingroup \@sanitize@url \@href}%
\providecommand \@href[1]{\@@startlink{#1}\@@href}%
\providecommand \@@href[1]{\endgroup#1\@@endlink}%
\providecommand \@sanitize@url [0]{\catcode `\\12\catcode `\$12\catcode
  `\&12\catcode `\#12\catcode `\^12\catcode `\_12\catcode `\%12\relax}%
\providecommand \@@startlink[1]{}%
\providecommand \@@endlink[0]{}%
\providecommand \url  [0]{\begingroup\@sanitize@url \@url }%
\providecommand \@url [1]{\endgroup\@href {#1}{\urlprefix }}%
\providecommand \urlprefix  [0]{URL }%
\providecommand \Eprint [0]{\href }%
\providecommand \doibase [0]{http://dx.doi.org/}%
\providecommand \selectlanguage [0]{\@gobble}%
\providecommand \bibinfo  [0]{\@secondoftwo}%
\providecommand \bibfield  [0]{\@secondoftwo}%
\providecommand \translation [1]{[#1]}%
\providecommand \BibitemOpen [0]{}%
\providecommand \bibitemStop [0]{}%
\providecommand \bibitemNoStop [0]{.\EOS\space}%
\providecommand \EOS [0]{\spacefactor3000\relax}%
\providecommand \BibitemShut  [1]{\csname bibitem#1\endcsname}%
\let\auto@bib@innerbib\@empty
\bibitem [{\citenamefont {Garfinkel}\ \emph {et~al.}(1997)\citenamefont
  {Garfinkel}, \citenamefont {Chen}, \citenamefont {Walter}, \citenamefont
  {Karagueuzian}, \citenamefont {Kogan}, \citenamefont {Evans}, \citenamefont
  {Karpoukhin}, \citenamefont {Hwang}, \citenamefont {Uchida}, \citenamefont
  {Gotoh}, \citenamefont {Nwasokwa}, \citenamefont {Sager},\ and\ \citenamefont
  {Weiss}}]{Garfinkel97}%
  \BibitemOpen
  \bibfield  {author} {\bibinfo {author} {\bibfnamefont {A.}~\bibnamefont
  {Garfinkel}}, \bibinfo {author} {\bibfnamefont {P.-S.}\ \bibnamefont {Chen}},
  \bibinfo {author} {\bibfnamefont {D.~O.}\ \bibnamefont {Walter}}, \bibinfo
  {author} {\bibfnamefont {H.~S.}\ \bibnamefont {Karagueuzian}}, \bibinfo
  {author} {\bibfnamefont {B.}~\bibnamefont {Kogan}}, \bibinfo {author}
  {\bibfnamefont {S.~J.}\ \bibnamefont {Evans}}, \bibinfo {author}
  {\bibfnamefont {M.}~\bibnamefont {Karpoukhin}}, \bibinfo {author}
  {\bibfnamefont {C.}~\bibnamefont {Hwang}}, \bibinfo {author} {\bibfnamefont
  {T.}~\bibnamefont {Uchida}}, \bibinfo {author} {\bibfnamefont
  {M.}~\bibnamefont {Gotoh}}, \bibinfo {author} {\bibfnamefont
  {O.}~\bibnamefont {Nwasokwa}}, \bibinfo {author} {\bibfnamefont
  {P.}~\bibnamefont {Sager}}, \ and\ \bibinfo {author} {\bibfnamefont {J.~N.}\
  \bibnamefont {Weiss}},\ }\bibfield  {title} {\enquote {\bibinfo {title}
  {Quasiperiodicity and chaos in cardiac fibrillation},}\ }\href@noop {}
  {\bibfield  {journal} {\bibinfo  {journal} {J. of Clinical Investigation}\
  }\textbf {\bibinfo {volume} {99}},\ \bibinfo {pages} {305} (\bibinfo {year}
  {1997})}\BibitemShut {NoStop}%
\bibitem [{\citenamefont {Goldobin}\ and\ \citenamefont
  {Pikovsky}(2006)}]{Goldobin06}%
  \BibitemOpen
  \bibfield  {author} {\bibinfo {author} {\bibfnamefont {D.~S.}\ \bibnamefont
  {Goldobin}}\ and\ \bibinfo {author} {\bibfnamefont {A.}~\bibnamefont
  {Pikovsky}},\ }\bibfield  {title} {\enquote {\bibinfo {title}
  {Antireliability of noise-driven neurons},}\ }\href@noop {} {\bibfield
  {journal} {\bibinfo  {journal} {Phys. Rev. E}\ }\textbf {\bibinfo {volume}
  {73}},\ \bibinfo {pages} {061906} (\bibinfo {year} {2006})}\BibitemShut
  {NoStop}%
\bibitem [{\citenamefont {Brown}, \citenamefont {Chua},\ and\ \citenamefont
  {Popp}(1992)}]{Brown92}%
  \BibitemOpen
  \bibfield  {author} {\bibinfo {author} {\bibfnamefont {R.}~\bibnamefont
  {Brown}}, \bibinfo {author} {\bibfnamefont {L.}~\bibnamefont {Chua}}, \ and\
  \bibinfo {author} {\bibfnamefont {B.}~\bibnamefont {Popp}},\ }\bibfield
  {title} {\enquote {\bibinfo {title} {Is sensitive dependence on initial
  conditions nature's sensory device?}}\ }\href@noop {} {\bibfield  {journal}
  {\bibinfo  {journal} {International J. of Bifurcation and Chaos}\ }\textbf
  {\bibinfo {volume} {2}},\ \bibinfo {pages} {193} (\bibinfo {year}
  {1992})}\BibitemShut {NoStop}%
\bibitem [{\citenamefont {Neiman}\ \emph {et~al.}(2011)\citenamefont {Neiman},
  \citenamefont {Dierkes}, \citenamefont {Lindner}, \citenamefont {Han},\ and\
  \citenamefont {Shilnikov}}]{Neiman11}%
  \BibitemOpen
  \bibfield  {author} {\bibinfo {author} {\bibfnamefont {A.~B.}\ \bibnamefont
  {Neiman}}, \bibinfo {author} {\bibfnamefont {K.}~\bibnamefont {Dierkes}},
  \bibinfo {author} {\bibfnamefont {B.}~\bibnamefont {Lindner}}, \bibinfo
  {author} {\bibfnamefont {L.}~\bibnamefont {Han}}, \ and\ \bibinfo {author}
  {\bibfnamefont {A.~L.}\ \bibnamefont {Shilnikov}},\ }\bibfield  {title}
  {\enquote {\bibinfo {title} {Spontaneous voltage oscillations and response
  dynamics of a hodgkin-huxley type model of sensory hair cells},}\ }\href@noop
  {} {\bibfield  {journal} {\bibinfo  {journal} {The J. of Math. Neurosci.}\
  }\textbf {\bibinfo {volume} {1}} (\bibinfo {year} {2011})}\BibitemShut
  {NoStop}%
\bibitem [{\citenamefont {Faber}\ and\ \citenamefont {Bozovic}()}]{Faber18b}%
  \BibitemOpen
  \bibfield  {author} {\bibinfo {author} {\bibfnamefont {J.}~\bibnamefont
  {Faber}}\ and\ \bibinfo {author} {\bibfnamefont {D.}~\bibnamefont
  {Bozovic}},\ }\bibfield  {title} {\enquote {\bibinfo {title} {Chaotic
  dynamics enhance the sensitivity of inner ear hair cells},}\ }\href@noop {}
  {\bibinfo  {journal} {[arXiv:1812.05156]}\ }\BibitemShut {NoStop}%
\bibitem [{\citenamefont {Hudspeth}(2014)}]{Hudspeth14}%
  \BibitemOpen
\bibfield  {journal} {  }\bibfield  {author} {\bibinfo {author} {\bibfnamefont
  {A.~J.}\ \bibnamefont {Hudspeth}},\ }\bibfield  {title} {\enquote {\bibinfo
  {title} {Integrating the active process of hair cells with cochlear
  function},}\ }\href@noop {} {\bibfield  {journal} {\bibinfo  {journal} {Nat.
  Rev. Neurosci.}\ }\textbf {\bibinfo {volume} {15}},\ \bibinfo {pages} {600}
  (\bibinfo {year} {2014})}\BibitemShut {NoStop}%
\bibitem [{\citenamefont {Leshowitz}(1971)}]{Leshowitz71}%
  \BibitemOpen
  \bibfield  {author} {\bibinfo {author} {\bibfnamefont {B.}~\bibnamefont
  {Leshowitz}},\ }\bibfield  {title} {\enquote {\bibinfo {title} {Measurement
  of the two‐click threshold},}\ }\href@noop {} {\bibfield  {journal}
  {\bibinfo  {journal} {The J. of the Acoustical Society of America}\ }\textbf
  {\bibinfo {volume} {49}},\ \bibinfo {pages} {462} (\bibinfo {year}
  {1971})}\BibitemShut {NoStop}%
\bibitem [{\citenamefont {Reichenbach}\ and\ \citenamefont
  {J.Hudspeth}(2014)}]{Reichenbach14}%
  \BibitemOpen
  \bibfield  {author} {\bibinfo {author} {\bibfnamefont {T.}~\bibnamefont
  {Reichenbach}}\ and\ \bibinfo {author} {\bibfnamefont {A.}~\bibnamefont
  {J.Hudspeth}},\ }\bibfield  {title} {\enquote {\bibinfo {title} {The physics
  of hearing: fluid mechanics and the active process of the inner ear},}\
  }\href@noop {} {\bibfield  {journal} {\bibinfo  {journal} {Reports on
  Progress in Physics}\ }\textbf {\bibinfo {volume} {77}} (\bibinfo {year}
  {2014})}\BibitemShut {NoStop}%
\bibitem [{\citenamefont {LeMasurier}\ and\ \citenamefont
  {Gillespie}(2005)}]{LeMasurier05}%
  \BibitemOpen
  \bibfield  {author} {\bibinfo {author} {\bibfnamefont {M.}~\bibnamefont
  {LeMasurier}}\ and\ \bibinfo {author} {\bibfnamefont {P.~G.}\ \bibnamefont
  {Gillespie}},\ }\bibfield  {title} {\enquote {\bibinfo {title} {Hair-cell
  mechanotransduction and cochlear amplification},}\ }\href@noop {} {\bibfield
  {journal} {\bibinfo  {journal} {Neuron}\ }\textbf {\bibinfo {volume} {48}},\
  \bibinfo {pages} {403} (\bibinfo {year} {2005})}\BibitemShut {NoStop}%
\bibitem [{\citenamefont {Martin}\ \emph {et~al.}(2003)\citenamefont {Martin},
  \citenamefont {Bozovic}, \citenamefont {Choe},\ and\ \citenamefont
  {Hudspeth}}]{Martin03}%
  \BibitemOpen
  \bibfield  {author} {\bibinfo {author} {\bibfnamefont {P.}~\bibnamefont
  {Martin}}, \bibinfo {author} {\bibfnamefont {D.}~\bibnamefont {Bozovic}},
  \bibinfo {author} {\bibfnamefont {Y.}~\bibnamefont {Choe}}, \ and\ \bibinfo
  {author} {\bibfnamefont {A.~J.}\ \bibnamefont {Hudspeth}},\ }\bibfield
  {title} {\enquote {\bibinfo {title} {Spontaneous oscillation by hair bundles
  of the bullfrog's sacculus},}\ }\href@noop {} {\bibfield  {journal} {\bibinfo
   {journal} {The J. of Neurosci.}\ }\textbf {\bibinfo {volume} {23}},\
  \bibinfo {pages} {4533} (\bibinfo {year} {2003})}\BibitemShut {NoStop}%
\bibitem [{\citenamefont {Martin}, \citenamefont {Hudspeth},\ and\
  \citenamefont {J{\"{u}}licher}(2001)}]{Martin01}%
  \BibitemOpen
  \bibfield  {author} {\bibinfo {author} {\bibfnamefont {P.}~\bibnamefont
  {Martin}}, \bibinfo {author} {\bibfnamefont {A.~J.}\ \bibnamefont
  {Hudspeth}}, \ and\ \bibinfo {author} {\bibfnamefont {F.}~\bibnamefont
  {J{\"{u}}licher}},\ }\bibfield  {title} {\enquote {\bibinfo {title}
  {Comparison of a hair bundle's spontaneous oscillations with its response to
  mechanical stimulation reveals the underlying active process},}\ }\href@noop
  {} {\bibfield  {journal} {\bibinfo  {journal} {Proc. Natl. Acad. Sci.}\
  }\textbf {\bibinfo {volume} {98}},\ \bibinfo {pages} {14380} (\bibinfo {year}
  {2001})}\BibitemShut {NoStop}%
\bibitem [{\citenamefont {Hudspeth}(2008)}]{Hudspeth08}%
  \BibitemOpen
  \bibfield  {author} {\bibinfo {author} {\bibfnamefont {A.~J.}\ \bibnamefont
  {Hudspeth}},\ }\bibfield  {title} {\enquote {\bibinfo {title} {Making an
  effort to listen: Mechanical amplification in the ear},}\ }\href@noop {}
  {\bibfield  {journal} {\bibinfo  {journal} {Neuron}\ }\textbf {\bibinfo
  {volume} {59}},\ \bibinfo {pages} {530} (\bibinfo {year} {2008})}\BibitemShut
  {NoStop}%
\bibitem [{\citenamefont {Faber}\ and\ \citenamefont
  {Bozovic}(2018)}]{Faber18a}%
  \BibitemOpen
  \bibfield  {author} {\bibinfo {author} {\bibfnamefont {J.}~\bibnamefont
  {Faber}}\ and\ \bibinfo {author} {\bibfnamefont {D.}~\bibnamefont
  {Bozovic}},\ }\bibfield  {title} {\enquote {\bibinfo {title} {Chaotic
  dynamics of inner ear hair cells},}\ }\href@noop {} {\bibfield  {journal}
  {\bibinfo  {journal} {Sci. Reports}\ }\textbf {\bibinfo {volume} {8}}
  (\bibinfo {year} {2018})}\BibitemShut {NoStop}%
\bibitem [{\citenamefont {Marsden}\ and\ \citenamefont
  {McCracken}(1976)}]{Marsden76}%
  \BibitemOpen
  \bibfield  {author} {\bibinfo {author} {\bibfnamefont {J.~E.}\ \bibnamefont
  {Marsden}}\ and\ \bibinfo {author} {\bibfnamefont {M.}~\bibnamefont
  {McCracken}},\ }\href@noop {} {\emph {\bibinfo {title} {The Hopf Bifurcation
  and its Appications}}}\ (\bibinfo  {publisher} {Springer-Verlag New York},\
  \bibinfo {year} {1976})\BibitemShut {NoStop}%
\bibitem [{\citenamefont {Egu{\'{i}}luz}\ \emph {et~al.}(2000)\citenamefont
  {Egu{\'{i}}luz}, \citenamefont {Ospeck}, \citenamefont {Choe}, \citenamefont
  {Hudspeth},\ and\ \citenamefont {Magnasco}}]{Eguiluz00}%
  \BibitemOpen
  \bibfield  {author} {\bibinfo {author} {\bibfnamefont {V.~M.}\ \bibnamefont
  {Egu{\'{i}}luz}}, \bibinfo {author} {\bibfnamefont {M.}~\bibnamefont
  {Ospeck}}, \bibinfo {author} {\bibfnamefont {Y.}~\bibnamefont {Choe}},
  \bibinfo {author} {\bibfnamefont {A.~J.}\ \bibnamefont {Hudspeth}}, \ and\
  \bibinfo {author} {\bibfnamefont {M.~O.}\ \bibnamefont {Magnasco}},\
  }\bibfield  {title} {\enquote {\bibinfo {title} {Essential nonlinearities in
  hearing},}\ }\href@noop {} {\bibfield  {journal} {\bibinfo  {journal} {Phys.
  Rev. Lett.}\ }\textbf {\bibinfo {volume} {84}},\ \bibinfo {pages} {5232}
  (\bibinfo {year} {2000})}\BibitemShut {NoStop}%
\bibitem [{\citenamefont {Kern}\ and\ \citenamefont {Stoop}(2003)}]{Kern03}%
  \BibitemOpen
  \bibfield  {author} {\bibinfo {author} {\bibfnamefont {A.}~\bibnamefont
  {Kern}}\ and\ \bibinfo {author} {\bibfnamefont {R.}~\bibnamefont {Stoop}},\
  }\bibfield  {title} {\enquote {\bibinfo {title} {Essential role of couplings
  between hearing nonlinearities},}\ }\href@noop {} {\bibfield  {journal}
  {\bibinfo  {journal} {Phys. Rev. Lett.}\ }\textbf {\bibinfo {volume} {91}},\
  \bibinfo {pages} {128101} (\bibinfo {year} {2003})}\BibitemShut {NoStop}%
\bibitem [{\citenamefont {Pikovsky}, \citenamefont {Rosenblum},\ and\
  \citenamefont {Kurths}(2001)}]{Pikovsky01}%
  \BibitemOpen
  \bibfield  {author} {\bibinfo {author} {\bibfnamefont {A.}~\bibnamefont
  {Pikovsky}}, \bibinfo {author} {\bibfnamefont {M.}~\bibnamefont {Rosenblum}},
  \ and\ \bibinfo {author} {\bibfnamefont {J.}~\bibnamefont {Kurths}},\
  }\href@noop {} {\emph {\bibinfo {title} {Synchronization: A Universal Concept
  in Nonlinear Sciences}}}\ (\bibinfo  {publisher} {Cambridge University
  Press},\ \bibinfo {year} {2001})\BibitemShut {NoStop}%
\bibitem [{\citenamefont {Benser}, \citenamefont {Marquis},\ and\ \citenamefont
  {Hudspeth}(1996)}]{Benser96}%
  \BibitemOpen
  \bibfield  {author} {\bibinfo {author} {\bibfnamefont {M.~E.}\ \bibnamefont
  {Benser}}, \bibinfo {author} {\bibfnamefont {R.~E.}\ \bibnamefont {Marquis}},
  \ and\ \bibinfo {author} {\bibfnamefont {A.~J.}\ \bibnamefont {Hudspeth}},\
  }\bibfield  {title} {\enquote {\bibinfo {title} {Rapid, active hair bundle
  movements in hair cells from the bullfrog’s sacculus},}\ }\href@noop {}
  {\bibfield  {journal} {\bibinfo  {journal} {J. of Neurosci.}\ }\textbf
  {\bibinfo {volume} {16}},\ \bibinfo {pages} {5629} (\bibinfo {year}
  {1996})}\BibitemShut {NoStop}%
\bibitem [{\citenamefont {Tinevez}, \citenamefont {J{\"{u}}licher},\ and\
  \citenamefont {Martin}(2007)}]{Tinevez07}%
  \BibitemOpen
  \bibfield  {author} {\bibinfo {author} {\bibfnamefont {J.-Y.}\ \bibnamefont
  {Tinevez}}, \bibinfo {author} {\bibfnamefont {F.}~\bibnamefont
  {J{\"{u}}licher}}, \ and\ \bibinfo {author} {\bibfnamefont {P.}~\bibnamefont
  {Martin}},\ }\bibfield  {title} {\enquote {\bibinfo {title} {Unifying the
  various incarnations of active hair-bundle motility by the vertebrate hair
  cell},}\ }\href@noop {} {\bibfield  {journal} {\bibinfo  {journal}
  {Biophysical J.}\ }\textbf {\bibinfo {volume} {93}},\ \bibinfo {pages} {4053}
  (\bibinfo {year} {2007})}\BibitemShut {NoStop}%
\bibitem [{\citenamefont {Zhou}\ and\ \citenamefont {Kurths}(2002)}]{Zhou02}%
  \BibitemOpen
  \bibfield  {author} {\bibinfo {author} {\bibfnamefont {C.}~\bibnamefont
  {Zhou}}\ and\ \bibinfo {author} {\bibfnamefont {J.}~\bibnamefont {Kurths}},\
  }\bibfield  {title} {\enquote {\bibinfo {title} {Noise-induced phase
  synchronization and synchronization transitions in chaotic oscillators},}\
  }\href@noop {} {\bibfield  {journal} {\bibinfo  {journal} {Phys. Rev. Lett.}\
  }\textbf {\bibinfo {volume} {88}},\ \bibinfo {pages} {230602} (\bibinfo
  {year} {2002})}\BibitemShut {NoStop}%
\bibitem [{\citenamefont {Strogatz}(1994)}]{Strogatz94}%
  \BibitemOpen
  \bibfield  {author} {\bibinfo {author} {\bibfnamefont {S.~H.}\ \bibnamefont
  {Strogatz}},\ }\href@noop {} {\emph {\bibinfo {title} {Nonlinear Dynamics and
  Chaos}}}\ (\bibinfo  {publisher} {Addison-Wesley},\ \bibinfo {year}
  {1994})\BibitemShut {NoStop}%
\bibitem [{\citenamefont {Gonzalez}\ and\ \citenamefont
  {Piro}(1983)}]{Gonzalez83}%
  \BibitemOpen
  \bibfield  {author} {\bibinfo {author} {\bibfnamefont {D.~L.}\ \bibnamefont
  {Gonzalez}}\ and\ \bibinfo {author} {\bibfnamefont {O.}~\bibnamefont
  {Piro}},\ }\bibfield  {title} {\enquote {\bibinfo {title} {Chaos in a
  nonlinear driven oscillator with exact solution},}\ }\href@noop {} {\bibfield
   {journal} {\bibinfo  {journal} {Phys. Rev. Lett.}\ }\textbf {\bibinfo
  {volume} {50}},\ \bibinfo {pages} {870} (\bibinfo {year} {1983})}\BibitemShut
  {NoStop}%
\bibitem [{\citenamefont {Patuzzi}\ and\ \citenamefont
  {Thompson}(1991)}]{Patuzzi91}%
  \BibitemOpen
  \bibfield  {author} {\bibinfo {author} {\bibfnamefont {R.~B.}\ \bibnamefont
  {Patuzzi}}\ and\ \bibinfo {author} {\bibfnamefont {M.~L.}\ \bibnamefont
  {Thompson}},\ }\bibfield  {title} {\enquote {\bibinfo {title} {Cochlear
  efferent neurones and protection against acoustic trauma: Protection of outer
  hair cell receptor current and interanimal variability},}\ }\href@noop {}
  {\bibfield  {journal} {\bibinfo  {journal} {Hearing Research}\ }\textbf
  {\bibinfo {volume} {54}},\ \bibinfo {pages} {45} (\bibinfo {year}
  {1991})}\BibitemShut {NoStop}%
\end{thebibliography}%

\end{document}